# Performance Studies of the p-spray/p-stop implanted Si Sensors for the SiD Detector

*Pooja Saxena [1], Kirti Ranjan, Ashutosh Bhardwaj, R. K. Shivpuri &*
*Satyaki Bhattacharya*

*(1) Centre for Detector & Related Software Technology, Department of Physics & Astrophysics, University of Delhi, Delhi-110007, India, psaxena@physics.du.ac.in*

**Abstract**

Silicon Detector (SiD) is one of the proposed detector for future $e^+e^-$ Linear colliders, like International Linear Collider (ILC). The estimated neutron background for ILC is around 1 - 1.6 x $10^{10}$ 1-MeV equivalent neutrons $cm^{-2}$ $year^{-1}$ for the Si micro strip sensors to be used in the innermost vertex detector. The $p^+n^-n^+$ double-sided Si strip sensors are supposed to be used as position sensitive sensors for SiD. On the $n^+n^-$ side of these sensors, shorting due to electron accumulation leads to uniform spreading of signal over all the $n^+$ strips. Hence inter-strip isolation becomes one of the major technological challenges. One of the attractive methods to achieve the inter-strip isolation is the use of uniform p-type implant on the silicon surface (p-spray). Another alternative is the use of floating p-type implants that surround the n-strips (p-stop). However, the high electric fields at the edge of the p-spray/p-stop have been shown to induce pre-breakdown micro-discharge. An optimization of the implant dose profile of the p-spray and p-stop is required to achieve good electrical isolation while ensuring satisfactory breakdown performance of the Si sensors. In the present work, we report the preliminary results of simulation study performed on the $n^+n^-$ Si sensors, equipped with p-spray and p-stops, using SILVACO tools.

## Introduction

The $p^+n^-n^+$ double-sided Si strip sensors are planned to be used as position sensitive sensors for proposed SiD experiment for the future $e^+e^-$ colliders. One of the major technological challenge of the $n^+n^-$ side of the Si strip sensors is to achieve a good isolation between the adjacent $n^+$ strips during their full life span. The fixed positive oxide charge density at the Si/SiO$_2$ interface results in the creation of electron conduction layer between the two adjacent $n^+$ strips, which results in the degradation of inter-strip isolation. In non-irradiated sensors, fixed positive oxide charge density ($Q_F$) (~$2x10^{11}$ $cm^{-2}$) is present due to fabrication process. Si detectors in the proposed lepton colliders will face substantial neutron fluence, for example innermost vertex of the proposed International Linear Collider (ILC) will face substantial neutron background of around 1-1.6 x $10^{10}$ 1-MeV equivalent neutrons $cm^{-2}$ $year^{-1}$ [1]. Irradiation leads to so called bulk damage & surface damage. Bulk damage will result in the decrease in charge collection efficiency due to charge carrier trapping & an increase in the leakage current due to an increase of generation/recombination centres. The surface damage causes higher values of the fixed positive oxide charges density (as high as 2x $10^{12}$ $cm^{-2}$) [2] & interface trap density at the Si-SiO$_2$ interface, which further degrades the inter strip isolation. Isolation techniques, which are commonly used to ensure good isolation, are p-spray [3], p-stop [3, 4] & combined use of both [5]. The p-spray method consists of having uniform $p^+$ layer beneath the Si/SiO$_2$ interface. Another alternative method, commonly known as p-stop method involves floating p$^+$ strip in between two adjacent n$^+$ strips. However, the high electric fields at the edge of the p-spray/p-stop techniques have been shown to induce early-breakdown [6]. The p-stop technology has the drawback of adding a mask level to the fabrication process that increases its complexity and cost over the p-spray technique. An optimization of the implant dose profile of the p-spray is required to achieve good electrical isolation while ensuring satisfactory breakdown performance of the Si strip sensors. In this work, we have studied the effect of the p-spray isolation technique on the n$^+$n$^-$ side of the double sided p$^+$n$^-$n$^+$ Si strip sensors.

## Simulation Technique

Schematic of the n$^+$n$^-$ structure of 80 x 320 μm$^2$ (strip width=18 μm) rectangular cell is shown in figure 1.

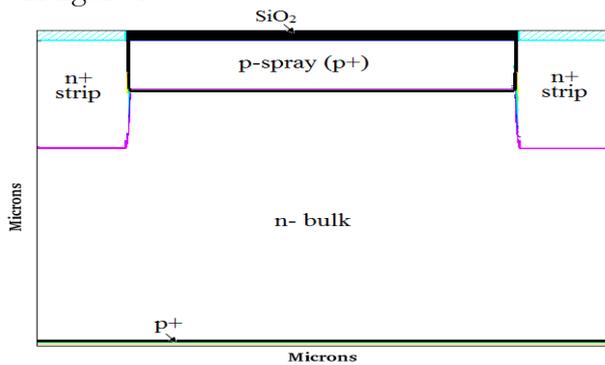

***Fig. 1:*** *Design structure of simulated n$^+$n$^-$ sensor (not to scale).*

The n-type wafer with uniform doping concentration (N$_B$) of 8.3x10$^{11}$ cm$^{-3}$ (resistivity of ~ 7.5 KΩ cm), junction depth (X$_j$) of 1.0 μm and oxide thickness (t$_{OX}$) of 1.0 μm, unless otherwise specified, is assumed. All n$^+$ implants are approximated by assuming a Gaussian profile with a peak concentration of 1x10$^{19}$ cm$^{-3}$ at the surface. It is assumed that the lateral junction width at the curvature of the implanted n$^+$ regions is equal to 0.7 times the vertical junction depth. Half of the cross section of the main n$^+$ strip is simulated because of the symmetric nature of the device. Three value of the peak dopant concentration of the p-spray (N$_P$) are considered in this study: Chosen values of N$_P$ are: 4.0x10$^{16}$ cm$^{-3}$ (low-dose p-spray), 12x10$^{16}$ cm$^{-3}$ (medium-dose p-spray) and 24x10$^{16}$ cm$^{-3}$ (high-dose p-spray). Depletion is attained by negatively biasing the backside contact, keeping the n$^+$ implant at the ground potential and p-spray as floating. The above mentioned device structure is used to study the AC characteristics & breakdown analysis using two-dimensional device simulation program, ATLAS version 5.15.32.R [7]. ATLAS solves Poisson's equation, continuity equation, energy-balance equation and the lattice heat equation for holes and electrons. To avoid the ambiguity in the ionization integral, ATLAS includes generated carriers due to impact ionization described by GRANT [7] directly in the solution of the device equation in a self-consistent manner. Reflecting Neumann conditions are imposed at the outer edges of the structure. The ohmic contacts at the main n$^+$ strip and backside contact are implemented using Dirichlet boundary conditions.

## Results

Depending on the dose of the p-spray implant, and for a given bias and Q$_F$, surface of the Si sensors just beneath the Si/SiO$_2$ interface can operate in any of the three modes as shown in figure 2, viz; (a) conduction mode, corresponding to electron accumulation layer, for N$_p$ < electron concentration (n$_e$), (b) depletion mode for N$_p$ = n$_e$, or (c) inversion mode for N$_p$ > n$_e$. In order to ensure good isolation between adjacent n$^+$ strips, Si sensors are needed to be operated in the depletion or inversion mode.

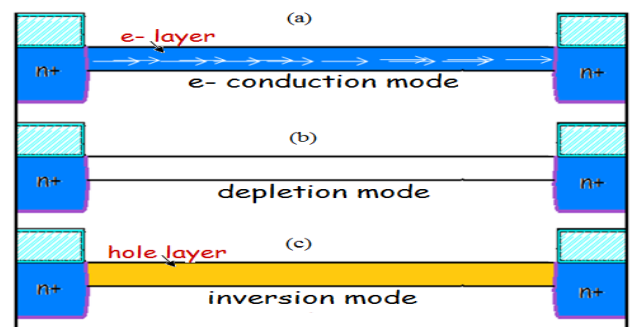

***Fig. 2:*** *Schematic showing isolation characteristics.*

Isolation characteristics of the n$^+$n$^-$ Si strip sensors are studied using inter-strip capacitance (C$_{int}$) and inter-strip conductance (G$_{int}$). Figures 3 and 4 show the plots of G$_{int}$ & C$_{int}$ vs. Q$_F$ for different doses of the p-spray at V$_{bias}$=-100 V respectively. For comparison purpose, we have also shown the results for Si sensor without p-spray.

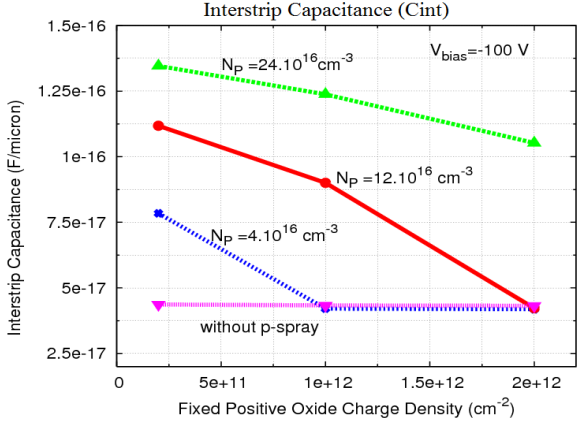

**Fig. 3:** $C_{int}$ vs $Q_F$ for p-spray & sensors without p-spray sensors.

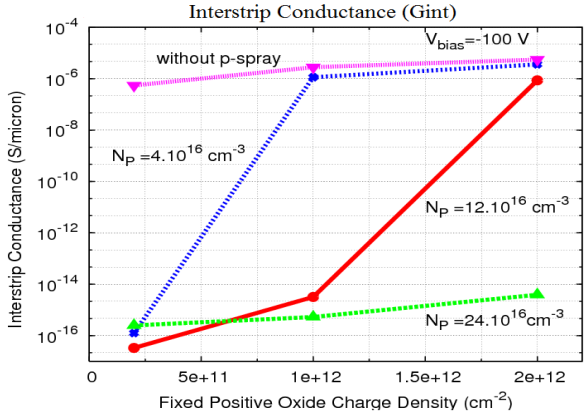

**Fig. 4:** $G_{int}$ vs $Q_F$ for p-spray & sensors without p-spray sensors.

At $Q_F = 2 \times 10^{11}$ cm$^{-2}$ (non-irradiated), for sensors without p-spray, $C_{int}$ is low but $G_{int}$ is high corresponding to poor isolation. Introduction of p-spray implant (all values of $N_P$) increases the $C_{int}$ (figure 3) but decreases the $G_{int}$ (figure 4), thus ensuring good inter-strip isolation in these sensors. Hence all $N_P$ values are effective in providing isolation at low $Q_F$. It can also be seen from these figures that for irradiated sensors, at $Q_F = 1 \times 10^{12}$ cm$^{-2}$, both medium and high p-spray dose can provide isolation. However, for $Q_F = 2 \times 10^{12}$ cm$^{-2}$, it turns out that only high p-spray dose implant can provide isolation. It is visible that $C_{int}$ decreases with increase in $Q_F$. Thus, it is clear that high value of $N_p$ is desirable in attaining good isolation between the n$^+$ strips in irradiated environment (for all $Q_F$ values). In order to understand it further, figure 5 shows the surface e$^-$ concentration plot for different values of p-spray dose with $Q_F = 1 \times 10^{12}$ cm$^{-2}$ at $V_{bias} = -100$V. The decrease in e$^-$ concentration with increase in p-spray dose further highlights the $G_{int}$ observation for $Q_F = 1 \times 10^{12}$ cm$^{-2}$.

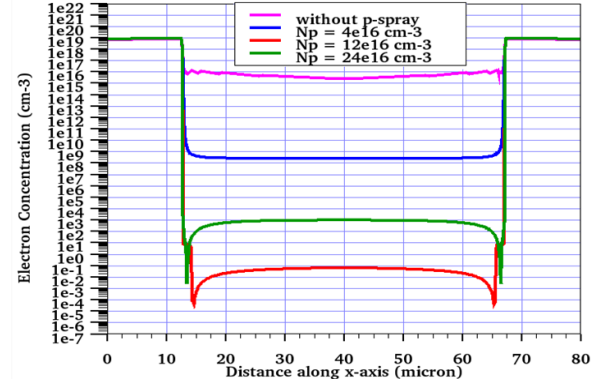

**Fig. 5:** e$^-$ concentration along x-axis (0.1μm below Si/SiO$_2$ interface) for p-spray & without p-spray sensors, $Q_F = 10 \times 10^{12}$ cm$^{-2}$ and $V_{bias} = -100V$.

For the sensors without p-spray, e$^-$ concentration between the two n$^+$ strips is found to be high enough to provide e$^-$ conduction layer, hence $G_{int}$ is found to have high value (figure 4). For low dose p-spray case, e$^-$ concentration is still very high & not providing adequate isolation, resulting in high $G_{int}$. For medium dose p-spray, e$^-$ concentration falls considerably and hence $G_{int}$ is low for medium dose p-spray (good isolation is achieved). Similarly with high dose p-spray, the effect is far more pronounced, and, as a consequence, isolation improves further. Hence e$^-$ (or hole) concentration plots help in better understanding of the isolation characteristics.

For sensors equipped with p-spray, highly doped n$^+$ implant comes in direct contact with p$^+$ implant (of p-spray), thus making this contact region very critical with respect to the breakdown characteristics. In these sensors, breakdown takes place due to avalanche caused by the high electric field around this region. Figure 6 shows the breakdown voltage ($V_{BD}$) vs. positive oxide charge density ($Q_F$) for different doses of p-spray implants. Again, for comparison purpose, we have also shown the results for the sensor without p-spray. For sensors without p-spray, breakdown voltage is ~2770 V at $Q_F = 2 \times 10^{11}$ cm$^{-2}$, which increases with increase in $Q_F$. For low dose p-spray, $V_{BD}$ is found to be less than for sensors without p-spray for all values of $Q_F$. Further increase in the value of $N_P$ results in further degradation of the $V_{BD}$. For high dose p-spray sensors, the $V_{BD}$ degrades drastically and is in the range of 150-400 V for all values of the $Q_F$ investigated.

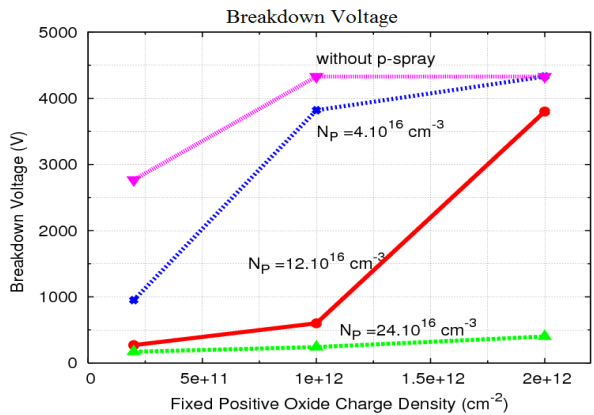

**Fig. 6:** *Breakdown voltage ($V_{BD}$) vs $Q_F$ for p-spray & without p-spray sensors.*

Corresponding surface electric field plot at $V_{bias}$= -100V & $Q_F$=1x10$^{12}$ cm$^{-2}$ as shown in figure 7, provides an insight to understand this observation. For the given $V_{bias}$, the peak surface electric field value is considerably higher for low dose p-spray implant than that for sensor without p-spray, hence the $V_{BD}$ is expected to occur at significantly lower values for sensors with p-spray. Increase in the p-spray dose leads to higher values of surface electric field, which in turn results in further decrease in $V_{BD}$.

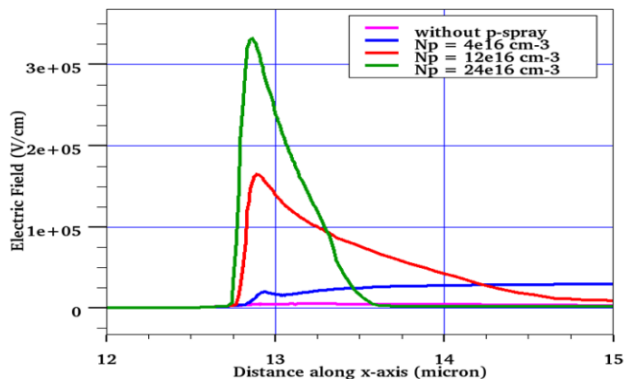

**Fig. 7:** *Electric field along x-axis (0.1μm below Si/SiO$_2$ interface) for p-spray & without p-spray sensors for $Q_F$=1x10$^{12}$cm$^{-2}$ & $V_{bias}$=-100V.*

To summarize, for achieving good electrical isolation of n$^+$n$^-$ Si strip sensor while ensuring satisfactory breakdown performance, an optimization of the implant dose profile of the p-spray is required.

## Conclusion

In this work, device simulation study is being performed on p-spray isolation technique for n$^+$n$^-$ Si strip sensors. An adequate dose of p-spray is required to compensate the electron conduction layer just beneath the Si/SiO$_2$ interface, providing the required isolation. Low dose p-spray sensor has high breakdown voltage for all values of $Q_F$ but does not provide isolation for high $Q_F$ values, while the high dose sensor, though provides better isolation for all $Q_F$ values, but, results in degradation of the breakdown performance. Electric field and e$^-$ (hole) concentration distributions can help in the understanding the breakdown & isolation characteristics of the n$^+$n$^-$ sensors.

## Acknowledgement

PS would like to thank CSIR for providing financial assistance.